\documentclass[11pt,twoside]{article}

\usepackage{asp2006}
\usepackage{epsf}
\usepackage{psfig}
\usepackage{lscape}

\begin{document}
\title{Creating Telescope Bibliographies Electronically -- Are We There Yet?}   
\author{Sarah Stevens-Rayburn}   
\affil{Space Telescope Science Institute, 3700 San Martin Drive,
Baltimore, MD 21218, USA}    

\author{Uta Grothkopf}   
\affil{European Southern Observatory, Karl-Schwarzschild-Str. 2, 85748
Garching, Germany}    

\begin{abstract}Observatory librarians traditionally have maintained databases of publications 
resulting from usage of their facilities.  In the era of electronic publication, the 
methodology of the creation of these databases has perforce changed as
well.  In this poster, we will compare a variety of methods for obtaining this information electronically 
and point out the advantages and shortcomings of each. 
\end{abstract}


\section{Introduction}
For those outside of an observatory, the question naturally arises 
as to why we would put all of this time and effort into documenting the numbers 
of papers written by users of a particular facility.  After all, our job as librarians is 
to gather, organize, and make available the scientific literature, not
to simply {\itshape count} it.  However, in these days of competing
demands for funding, those who  
supply that funding want to assure that they are putting their money
into the more  
productive projects.  The institutions for their part want to know
that those using  
their facilities are not merely collecting data, but are also putting
it to good use by  
publishing their results in the refereed literature and making their
data available  
for other scientists to use.  Finally, scientists compete for time on
heavily over-subscribed instruments.  Who actually gets to observe is
usually decided by a  
TAC -- time allocation committee and those providing the funding as well as the 
institutions want to know that the TAC is selecting the best of the proposals and 
``best'' is usually characterized as the most publishable, that which produces 
meaningful scientific output.

\section{Changing Methods for Changing Times}
Traditionally, bibliography compilers 
(usually librarians) reviewed all incoming journals for papers using a
particular  
facility.  This was a tedious, but relatively consistent task with little difference in 
methodology among different journals.  There was a steady pace of work that 
could be easily integrated into the library's traditional workflow.
In addition,  
once one became familiar with the standard abbreviations and paper
styles (i.e.,  
most journal papers have an ``observations'' or ``data used'' section),
the task went  
fairly quickly.  On the other hand, it was also easy to miss papers,
especially  
because authors are inconsistent in supplying the ``required''
acknowledgements. 

Given the advances in technology over the past decade or so, one may
now screen  
electronic tables of contents or individual journal websites to find
the relevant  
papers.  Additionally, there are a variety of aggregators of journal
metadata, such  
as the Astrophysics Data System (ADS), Web of Science, or Inspec, so one may 
search many journals simultaneously.  The advantages to electronic
searches are  
that, assuming the search is crafted sufficiently well, fewer papers
are missed and  
the selection of which papers need to be reviewed more carefully is
accelerated. 

\section{Current Studies}
The libraries at ESO, NRAO, and ST ScI each conducted 
separate experiments.  NRAO tracked the source of additions to their bibliography 
for five months.  During that time, 92\% of the papers added came from electronic 
tables of contents (TOCs), 6.5\% from author notifications or checking the print 
journals on arrival, and 1.5\% from ADS alerts.  At ST ScI,  emailed TOCs were 
searched for relevant articles and then rechecked when the paper
versions arrived.   
We then reversed the procedure and searched the paper first, followed after a 
suitable delay searching the electronic version.  This was done for a
five-month  
period at the end of which we performed a ``canned'' search in the ADS and 
compared the results against the bibliography.  The comparison showed 66.8\% of 
a total of 522 papers were found in both the ADS and the bibliography; 2.6\% 
were in the ADS but not in the bibliography, and 30.6\% in the bibliography but 
not the ADS.  There were also ten false drops from the ADS plus seven 
unrefereed papers returned as refereed.  ESO checked their traditional
acquisition  
method against the ADS for all of 2005 and found 62.4\%  (of 585 papers) in 
common, 36.9\% not in the ADS search, and less than 1\% in the ADS that the 
traditional method had missed.  ESO also noted 108 false drops in the ADS 
search, including 32 unrefereed conference papers cited as refereed by the ADS.
Of the remaining false drops, 47 could be traced to ambiguous acronyms and/or 
search terms, a problem endemic to automated searches unfortunately.

ESO then did a further study on a sample of 40 papers that did not
turn up in the  
ADS search and learned that the mention of ESO facilities had been in the full 
text of the articles for 42.5\% and that the ESO footnote did not make
it into the  
ADS indexing for 45\% of the papers.

\section{Results and General Observations}
To put it succinctly, we are not there yet.  
Each journal has its own unique methodology for searching content, some more 
useful for this kind of task than others.  For instance, we rated the
{\itshape Astrophysical  
Journal} and {\itshape Monthly Notices of the Royal Astronomical
Society} as the most  
``user-friendly'' since both allow for complex searches in the full-text
of individual  
issues. {\itshape MN} also highlights the search terms within the
paper, speeding the process 
of deciding whether a specific paper should be included or not. At the
opposite  
end of the spectrum is {\itshape Astronomy \& Astrophysics} whose
full-text searches are  
currently limited to 100 or fewer hits, and the {\itshape Publications of the
Astronomical  
Society of Japan}, whose HTML articles are broken into sections so one cannot 
search the whole article at one time.  (We should point out that when
this study  
was first done, {\itshape A\&A} presented their results with oldest
first, so that for an  
observatory with many papers,  one would never get to the current ones. They 
have now reversed the presentation so the latest are retrieved first.)
In addition to  
these very user-unfriendly sites, there are those that are merely
frustrating and  
time-consuming because they allow for complex searching, but not at the issue 
level or those whose complex searches are limited to the metadata;
i.e., searching  
only the title, abstract, and keywords.  At the end of this article is
a table that summarizes the searching advantages and limitations of
the major journals in the  field.
 
\section{The Astrophysics Data System}
Anyone familiar with astronomical 
bibliography is aware of the huge impact the ADS has had on the way 
astronomers (and astronomy librarians) do research and will naturally be 
wondering why that system isn't used exclusively for tasks such as this.  
Unfortunately, as the numbers in section 3 demonstrate, searching
just the ADS  
has severe limitations.  These include the fact that the ADS has no full-text 
searching of current journals.  One also cannot limit one's search to
a specific  
issue but even more problematic for this kind of work, we learned that
the more  
complex the search expression, the more likely the ADS would deliver
irrelevant  
hits, each of which has to be examined separately.  There is little
doubt that as  
electronic publishing matures (one must remember that the first e-journal in 
astronomy appeared as recently as 1996) and as searching techniques
advance that 
the ADS will become increasingly useful for compiling telescope
bibliographies,  
especially if one can be content with a completeness rate of 85-90\%
rather than the ideal 100\%.

\section{Conclusions}
In summary, we can say that e-screening for papers from a 
particular observatory works best when few hits are expected, the
journal has full-text searching capabilities, the title in question has a large number
of papers, but  
relatively few hits, and/or the number of search terms (that is
telescope/instrument  
names) are few.  Manual screening, on the other hand, still needs to
be used in  
cases where the journal doesn't allow for complex searching, the
journal is likely  
to have many relevant papers, and/or electronic searching is only
possible within  
the title, abstract, and keywords.

\begin{table}[!ht]
\caption{(1) Notes: ScienceDirect allows full text, complex searching by
subscribed (or non-subscribed) journals, but not for individual
titles.  Individual title searching is limited to  
title, abstract, and keywords.}
\smallskip
\begin{center}
{\small
\begin{tabular}{lcccc}
\tableline
\noalign{\smallskip}
Title & Full-text searching & Highlighting & Single issue searching &
HTML/pdf\\ 
\noalign{\smallskip}
\tableline
\noalign{\smallskip}
A\&A & Y $<$100 hits & N & N & both\\
\noalign{\smallskip}
ADS & N & Y & N & NA\\
\noalign{\smallskip}
AJ & Y & N & N & both\\
\noalign{\smallskip}
AN & Y & N & N & pdf\\
\noalign{\smallskip}
ApJ/ApJS & Y & N & Y & both\\
\noalign{\smallskip}
Astron. Lett. & Y & In hitlist only & N & pdf\\
\noalign{\smallskip}
GRL & Y & N & N & both\\
\noalign{\smallskip}
Icarus & Y (1) & Y & N & both\\
\noalign{\smallskip}
JGR & Y & N & N & both\\
\noalign{\smallskip}
MNRAS & Y & Y & Y & both\\
\noalign{\smallskip}
Nature/Science & Y & In hitlist only & N & both\\
\noalign{\smallskip}
New Astron. & Y (1) & Y & N & both\\
\noalign{\smallskip}
PASP & Y & N & N & both\\
\noalign{\smallskip}
\tableline
\end{tabular}
}
\end{center}
\end{table}


\acknowledgements 
The authors extend their thanks to Elizabeth Fraser (ST ScI), 
Angelika Treumann (ESO), and Marsha Bishop (NRAO) who supplied the necessary  
data on their searching experiences for this analysis.



\end{document}